\documentclass[twocolumn,preprintnumbers,amsmath,amssymb,article]{revtex4}

\usepackage{epsf}
\usepackage{rotating}

\usepackage{graphicx,epsfig}

\usepackage{dcolumn}
\usepackage{bm}
\usepackage{color}

\def\vmax{{v_{\rm max}}}

\def\vi{v_i}
\def\dix{\tilde{x}_i}
\def\diy{\tilde{y}_i}

\def\Di{(X_i, Y_i)}
\def\xit{x_i^t}
\def\yit{y_i^t}
\def\xito{x_i^{t+1}}
\def\yito{y_i^{t+1}}
\def\xjt{x_j^t}
\def\yjt{y_j^t}
\def\xitn{x_i^{t+n\delta t}}
\def\yitn{y_i^{t+n\delta t}}
\def\xitno{x_i^{t+(n+1)\delta t}}
\def\yitno{y_i^{t+(n+1)\delta t}}

\newcommand{\sgn}[1]{{\rm Sign}\left( #1\right)}

\newcommand{\tuple}[2]{\Big(#1, #2\Big)}

\newcommand{\req}[1]{Eq.~(\ref{#1})}

\newcommand{\fig}[1]{Fig.~\ref{#1}}

\newcommand{\cut}[1]{{}}
\newcommand{\etal}[1]{\emph{~et al.}}

\begin{document}

\preprint{}

\title[Title]
{The global benefit of randomness in individual routing on transportation networks}

\author{Tak Shing Tai and Chi Ho Yeung}
\email{chyeung@eduhk.hk}
\affiliation{Department of Science and Environmental Studies, The Education University of Hong Kong, Tai Po, Hong Kong}

\date{\today}

\begin{abstract}
By introducing a simple model based on two-dimensional cellular automata, we reveal the relationship between the routing strategies of individual vehicles and the global behavior of transportation networks. Specifically, we characterize the routing strategies by a single parameter called path-greediness, which corresponds to the tendency for individuals to travel via a shortest path to the destination. Remarkably, when vehicles tend to travel via the shortest path, a congested-flow state emerges between the conventional free-flow and congested states, where traffic flow increases slowly with vehicle density in the presence of congestion. We also found that a high individual tendency to travel via the shortest path does not necessarily shorten the average journey time, as the system may benefit from less greedy routing strategies in congested situations. Finally, we show that adaptive routing strategies outperform controlled strategies in the free-flow state, but not in the congested state, implying that controlled strategies may increase coordination among vehicles and are beneficial for suppressing traffic congestion.
\end{abstract}


\maketitle


\section{Introduction}

Traffic congestion is severe in many urban areas especially in cosmopolitan cities. Methods have been proposed to reduce traffic congestion and one of them is to optimize traffic flow as limited by the fixed infrastructure of roads. To reduce traffic congestion and to optimize traffic flow, route coordination is considered, as vehicles can make the best use of roads in the networks \cite{PhysRevLett.108.208701,Yeung13717}. In this case, some road users are diverted to longer paths to reduce the load on major routes, and thus to reduce traffic congestion. However, strategies to distribute users to longer paths correspond to a trade-off of traveling time between individual drivers and all drivers in the networks, i.e. a non-trivial balance between the social and the global optimum~\cite{youn2008price}. A study to understand and illustrate the impact of individual routing strategies on the macroscopic traffic condition is thus warranted.

Since experiments on real transportation networks would affect the daily movement of citizens, models are often employed to study the dynamics of vehicles. One seminal model for the study of vehicle interaction is the cellular automaton (CA) transportation model suggested by Nagel and Schreckenberg \cite{NSModel}. Many variants of the model exist, such as the particle hopping model which explains the traffic flow theory~\cite{PhysRevE.53.4655,CHOWDHURY2000199} and the model with two lanes and bidirectional traffic which reveals the interaction between vehicles~\cite{PhysRevE.57.2441}. Others have applied the CA model to understand vehicle dynamics, for instance, to study lane changing~\cite{DAGANZO2006396} and to explain traffic breakdowns and the characteristics of different transportation states~\cite{PhysRevE.84.046110}. Probabilistic CA models with single lanes are also introduced~\cite{PhysRevE.51.2939}. Similar to CA models, cell transmission models~\cite{JABARI2012156,  SUMALEE2011507} are applied to study transportation networks  and to reveal the relationship among metrics such as traveling time, average vehicle speed and traffic flow \cite{maerivoit2005_physicsReports}. All these models contribute to our increasing understanding of transportation networks.

Other than microscopic vehicle dynamics, there are studies to utilize models to connect microscopic factors with macroscopic behavior in transportation networks. For instance, a mathematical model was introduced to study human factors in driving behavior~\cite{VANLINT201863}; the so-called efficient hierarchical control strategies were developed to predict the traveling time and to identify traffic equilibrium~\cite{RAMEZANI20151}. In other studies, methods are introduced to predict traffic conditions. For instance, the functional principal component analysis was employed to forecast the traveling time between specific locations~\cite{ZHONG2017292}; a framework was introduced to evaluate the network performance with different pricing conditions~\cite{GU20181}; pricing systems were applied to study flow patterns~\cite{XU201882, ZHANG2018190}. The above studies showed that transportation networks are complex systems and each individual factor has important impacts on the system.

While transportation networks are dependent on many factors, one crucial factor is the routing strategy of individual drivers. To study the impact of routing strategies, two-lane or multiple-lane routing strategies have been studied and the benefit of these strategies are examined in different network scenarios~\cite{HUANG2017169,PhysRevE.82.066107,LE2017539,ZHAO201887}. Uncertainty in routing is another important factor, and was modeled in~\cite{JABARI2012156, SUBRAMANYAM2018296}; a choice model of stylized stochastic user equilibrium was employed to study two different routing scenarios to reduce traffic congestion~\cite{KOSTER2018137}. Other related studies include the introduction of a station sensitivity index to predict passenger flows~\cite{Silva5643}, and the application of stochastic differential equations to study traffic flow and phase transitions~\cite{SIQUEIRA20161}. These studies showed that routing strategies and random noise in routing have significant impacts on transportation networks.

In this paper, we will reveal the impact of individual preferences to travel through the shortest path on the global behavior of transportation networks. We will introduce a model based on cellular automata on square lattices, where individual drivers either route through the shorter path, or intentionally move to a random direction and travel through a longer path to the destination, according to a parameter called \emph{path-greediness}. Based on the model, we will reveal characteristics of the system such as phase transitions and the effectiveness of different routing strategies in reducing traffic congestion.

\section{Model}

Specifically, we consider a model of transportation networks on a two-dimensional $L\times L$ square lattice with periodic boundary condition, where each lattice site is labelled by the coordinate $(x,y)$, with $x,y=1,\dots, L$. There are a total of $N$ vehicles, labelled by $i=1, \dots, N$, running on the network by hoping between neighboring sites. Each site can be occupied by at most one vehicle. We denote the density of vehicles on the network to be $\rho=N/L^2$. Before the simulation starts, a pair of random origin and destination are drawn for each vehicle. Each vehicle $i$ then travels from its origin to its destination $\Di$; after it arrives at its destination, a new random destination is drawn and the vehicle starts to travel again. This transportation network is thus analogous to a system of 2-D lattice gas of which particles correspond to vehicles interacting with an individual time-varying potential. 

To model the vehicle movement, we denote the coordinate of vehicle $i$ at time $t$ to be $(\xit, \yit)$, its speed to be $\vi$ and the speed limit of the network to be $\vmax$, such that $\vi\le \vmax$ for all vehicle $i$. At each time step, one vehicle $i$ is picked randomly and moves to a new coordinate $(\xito, \yito)$ according to its adopted \emph{routing strategy}. The simulation repeats for $t$ steps, and various quantities of interest in the system such as the average vehicle speed and the number of completed journeys are measured.

\subsection{Routing strategy}
\label{sec_model}

To examine the impact of random noise on the choice of routes, we define a \emph{probabilistic routing strategy} where vehicles occasionally move in a direction away from the destination. For the sake of clarify in the discussion,
here we consider the case of $\vmax=1$ and discuss the cases with general $\vmax$ in Appendix~\ref{sec_vmax}. For a vehicle $i$ with coordinate $(\xit, \yit)$ at time $t$, its coordinate is updated by
\begin{align}
\label{eq_substep}
\tuple{\xito}{\yito} = \tuple{\xit}{\yit} + \eta_i^t\tuple{\Delta \xit}{\Delta \xit}.
\end{align}
where $\left(\Delta x_i^t, \Delta y_i^t\right)$ corresponds to the intended movement of vehicle $i$. Since each site can be occupied by at most one vehicle, the variable $\eta_i^t=0$ if the site of the next intended movement is already occupied by another vehicle, and otherwise $\eta_i^t=1$. In other words, $\eta_i^t$ is given by
\begin{align}
\eta_i^t =
\begin{cases}
0, &\hspace{-0.2cm}\mbox{if $\tuple{\xit}{\yit}\!+\!\tuple{\Delta x_i^t}{\Delta y_i^t}\!=\!\tuple{\xjt}{\yjt}, \exists j$}
\\
1, &\hspace{-0.2cm}\mbox{otherwise}
\end{cases}
\end{align}

Next, we describe how the intended movement $\left(\Delta x_i^t, \Delta y_i^t\right)$ is decided by vehicle $i$, since it is related to its routing strategy. To incorporate random noise in the path-selection process of drivers, we introduce a parameter $g$ which we call \emph{path-greediness}, whereas $0\le g\le 1$. With a probability $(1+g)/2$, a vehicle tends to move closer to its destination; with a probability $(1-g)/2$, the vehicle tends to move away from the destination. The path-greediness $g\ge 0$ as vehicles have a tendency to go to the destination; with $g=1$, vehicles always tend to move closer to the destination, and with $g=0$, vehicles move randomly. In other words, when vehicle $i$ has not yet arrived at neither the $x$- nor $y$-coordinate of its destination (i.e. $\xit\neq X_i$ and $\yit\neq Y_i$), we draw $(\Delta x_i^t, \Delta y_i^t)$ according to
\begin{align}
\label{eq_random1}
\left(\Delta x_i^t, \Delta y_i^t\right)=
\begin{cases}
\tuple{0}{\Delta \tilde{y}}, &\mbox{with a probability $\frac{1+g}{4}$},
\\
\tuple{\Delta \tilde{x}}{0}, &\mbox{with a probability $\frac{1+g}{4}$},
\\
\tuple{0}{-\Delta \tilde{y}}, &\mbox{with a probability $\frac{1-g}{4}$},
\\
\tuple{-\Delta \tilde{x}}{0}, &\mbox{with a probability $\frac{1-g}{4}$},
\\
\end{cases}
\nonumber\\
\end{align}
where $\Delta \tilde{y}$ and  $\Delta \tilde{x}$ are the greedy directions
\begin{align}
\label{eq_deltay}
\Delta \tilde{y}  = \sgn{Y_i-\yit} \sgn{\frac{L}{2}-|Y_i-\yit|}
\\
\label{eq_deltax}
\Delta \tilde{x}  = \sgn{X_i-\xit} \sgn{\frac{L}{2}-|X_i-\xit|}
\end{align}
with the sign function $\sgn{x}=1$ when $x\ge 0$, and otherwise $\sgn{x}=0$. The first sign function in Eqs. (\ref{eq_deltay}) and (\ref{eq_deltax}) determines respectively the $x$- and $y$-direction of the destination from the present location, while the second sign function reverses the movement direction as it is closer to move via the periodic boundary condition. 

On the other hand, when vehicle $i$ has arrived at either the $x$- or $y$-coordinate of its destination (i.e. $\xit=X_i$ and $\yit\neq Y_i$, or $\xit\neq X_i$ and $\yit= Y_i$), we assume that the vehicle moves in a direction towards  the destination with a probability $(1+3g)/4$. For instance, if $\xit= X_i$, the intended movement is given by 
\begin{align}
\label{eq_random1}
\left(\Delta \xit, \Delta \yit\right)=
\begin{cases}
\tuple{0}{\Delta \tilde{y}},&\mbox{with a probability $\frac{1+3g}{4}$},
\\
\tuple{1}{0},&\mbox{with a probability $\frac{1-g}{4}$},
\\
\tuple{0}{-\Delta \tilde{y}},&\mbox{with a probability $\frac{1-g}{4}$},
\\
\tuple{-1}{0}, &\mbox{with a probability $\frac{1-g}{4}$}.
\\
\end{cases}
\end{align}
Similarly, if $\yit= Y_i$, then
\begin{align}
\label{eq_random1}
\left(\Delta \xit, \Delta \yit\right)=
\begin{cases}
\tuple{\Delta\tilde{x}}{0}, &\mbox{with a probability $\frac{1+3g}{4}$},
\\
\tuple{0}{1}, &\mbox{with a probability $\frac{1-g}{4}$},
\\
\tuple{-\Delta\tilde{x}}{0}, &\mbox{with a probability $\frac{1-g}{4}$},
\\
\tuple{0}{-1}, &\mbox{with a probability $\frac{1-g}{4}$},
\\
\end{cases}
\end{align}
In cases with path-greediness $g=1$, vehicles are only directed towards their destinations and are always on a path with the shortest distance;  vehicles follow a random zig-zag trajectory to their destinations, which is one of the shortest paths. With path-greediness $g=0$, the movement of vehicles is completely random regardless of the location of their destinations.  

\subsection{Quantities of interest}

We assume that the simulation starts at time $t=0$ and equilibrate at time $t=T_e$, i.e. when all the quantities of interest become steady. We terminate the simulation at time $t=T$.  Several quantities will be measured to characterize the behavior of the system, which include the average vehicular speed $\bar{v}$, the arrival count per time step (i.e. the average number of vehicles arriving at their destination per time step), the average journey time and the average journey distance between origins and destinations.

Another crucial quantity of interest which characterizes the state of a transportation network is the \emph{average traffic flow} $f$, conventionally given by $f=\rho\bar{v}$~\cite{maerivoit2005_physicsReports}. In our model, the average traffic flow is equivalent to the average number of vehicle movements per time step per site, given by
\begin{align}
f = \frac{1}{(T-T_e)L^2} \sum_{t=T_e}^{T} \sum_{i=1}^{N} \eta_i^t
\end{align}
When $f$ increases with increasing vehicle density $\rho$, the system is in a free-flow state; on the other hand, when $f$ decreases with increasing $\rho$, the system is in a congested state~\cite{maerivoit2005_physicsReports}.

\section{Results}
 With our model, we employ computer simulations to reveal various macroscopic phenomena in transportation networks. Since our goal is to study the relationship between routing strategies and traffic congestion, we will focus on the system behavior and its dependence on the vehicle density $\rho$ and the path-greediness $g$, while keeping all other factors constant. In other words, we characterize routing strategies in our model by a single parameter $g$.

\subsection{The free-flow, the congested-flow and the congested states}
\label{sec_phase}

\begin{figure}
\includegraphics[ width=1\linewidth] {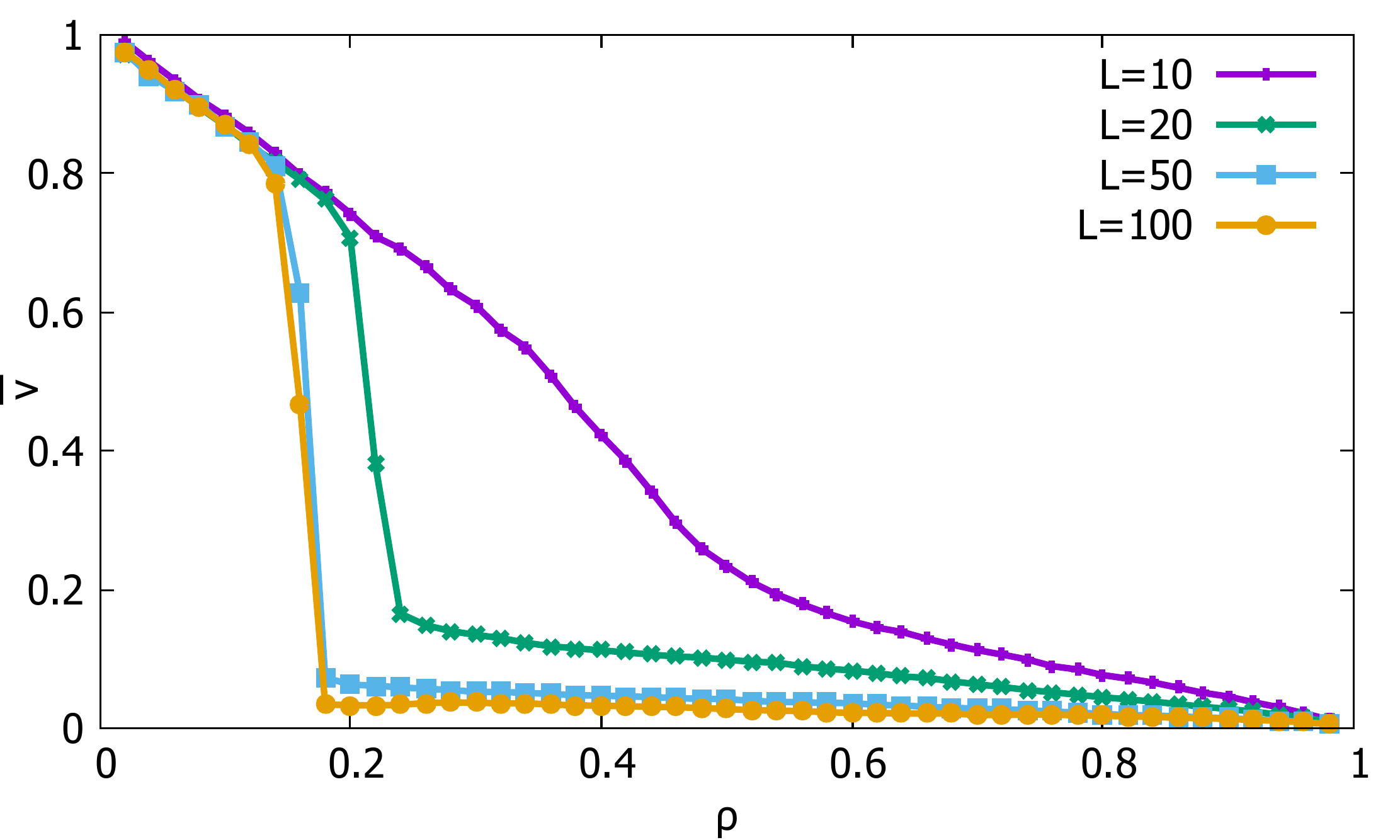}
\caption{
The simulation results of the average vehicle speed $\bar{v}$ as a function of $\rho$ with path-greediness $g=0.6$  on square lattices with various length $L$. The results are obtained with $T=3 \times 10^6$ and an equilibration time $T_e=2.5\times 10^6$, averaged over 1000 instances. 
}
\label{fig_diffL}
\end{figure}

As in other existing studies of transportation cellular automata, we first examine the dependence of the average vehicle speed $\bar{v}$ on vehicle density $\rho$.
As shown in \fig{fig_diffL}, $\bar{v}$ decreases abruptly at a specific value of $\rho_c$, as $\rho$ increases. We remark that the decrease of $\bar{v}$ is also observed in the fundamental model of cellular automata on a ring when $\rho$ increases beyond a threshold, but the decrease of $\bar{v}$ is more gentle in the ring topology~\cite{NSModel}. We also found that the decrease of $\bar{v}$ at  $\rho=\rho_c$ becomes more abrupt when $L$ increases, suggesting that a first order phase transition from a \emph{free-flow state} to a \emph{congested state} occurs at $\rho=\rho_c$.  These results imply that a first order phase transition may occur on transportation cellular automata with a dimension larger than one; additional results pointing to its relevance to the system dimension will be discussed later. The first order phase transition also implies that congestion may abruptly emerge in grid-like road networks given the slightest increase in vehicle density. Since the focus of this subsection is phase transition, we will further discuss the behavior of average speed $\bar{v}$ in the next subsection.

\begin{figure}
\includegraphics[ width=1\linewidth] {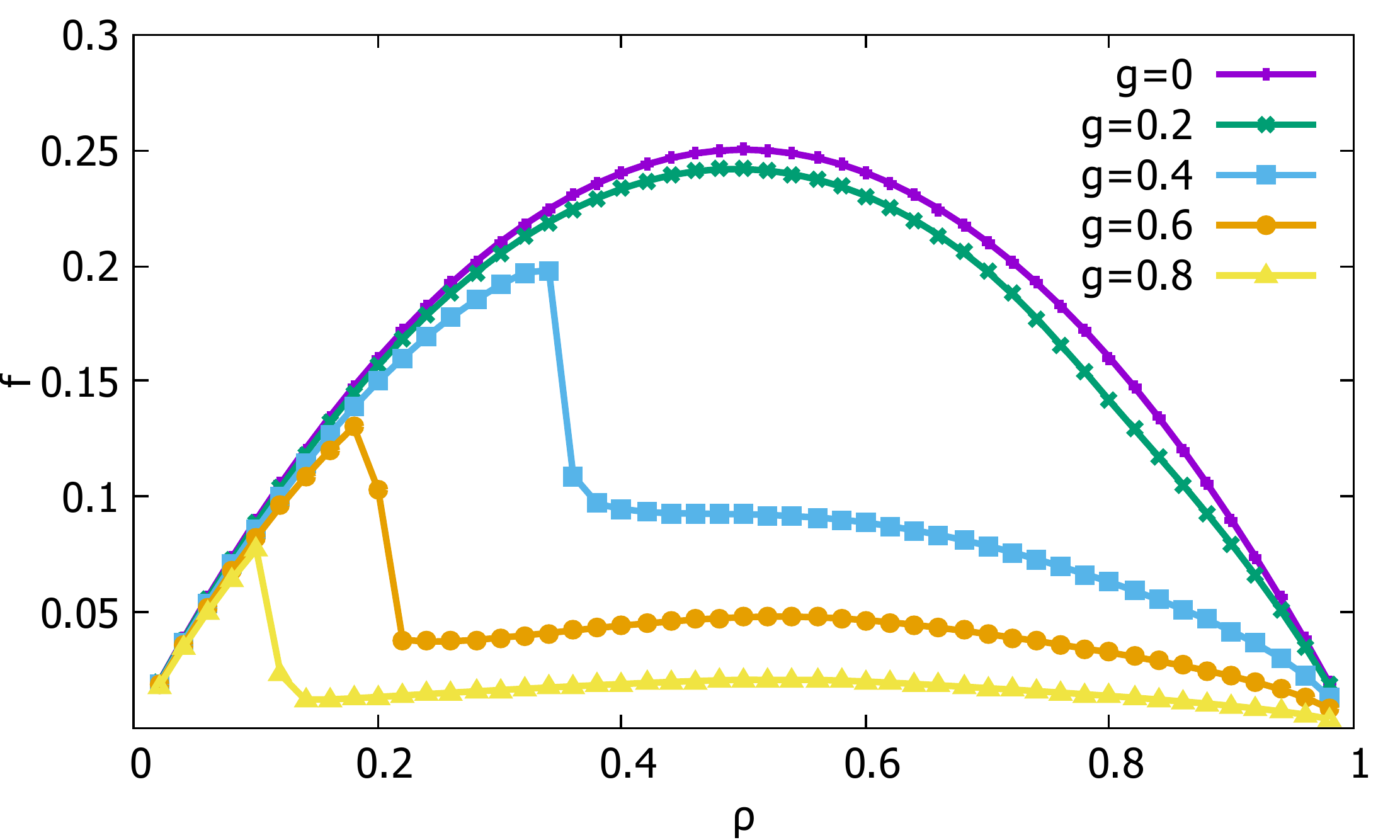}
\caption{
The simulation  results of the traffic flow $f$ as a function of $\rho$ with $L=20$ and various values of $g$. The results are obtained with $T=3 \times 10^6$ and an equilibration time $T_e=2.5\times 10^6$, averaged over 1000 instances.
}
\label{fig_flow}
\end{figure}

\begin{figure*}
\includegraphics[ trim=100 0 100 0,clip,width=0.3\linewidth] {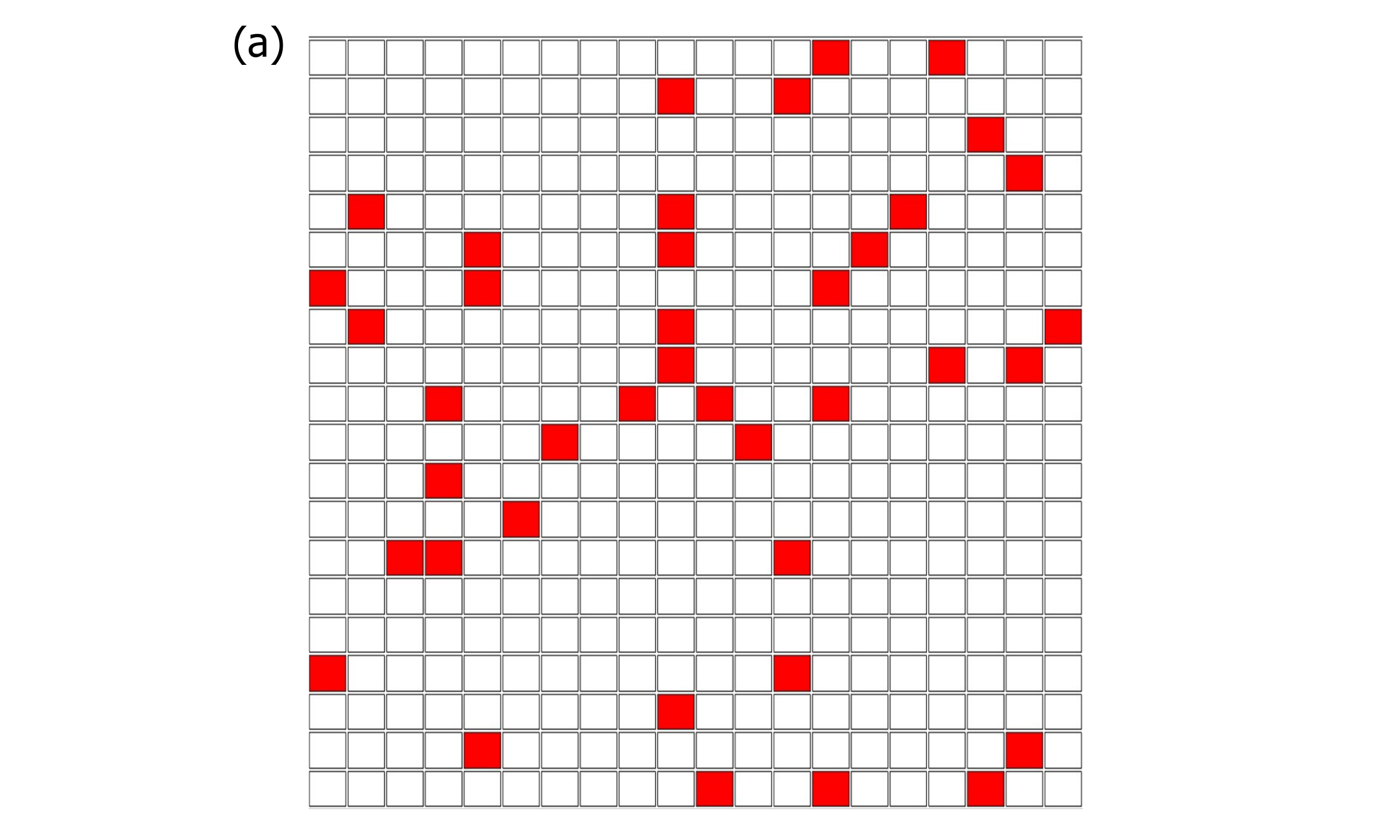}
\includegraphics[ trim=100 0 100 0,clip, width=0.3\linewidth] {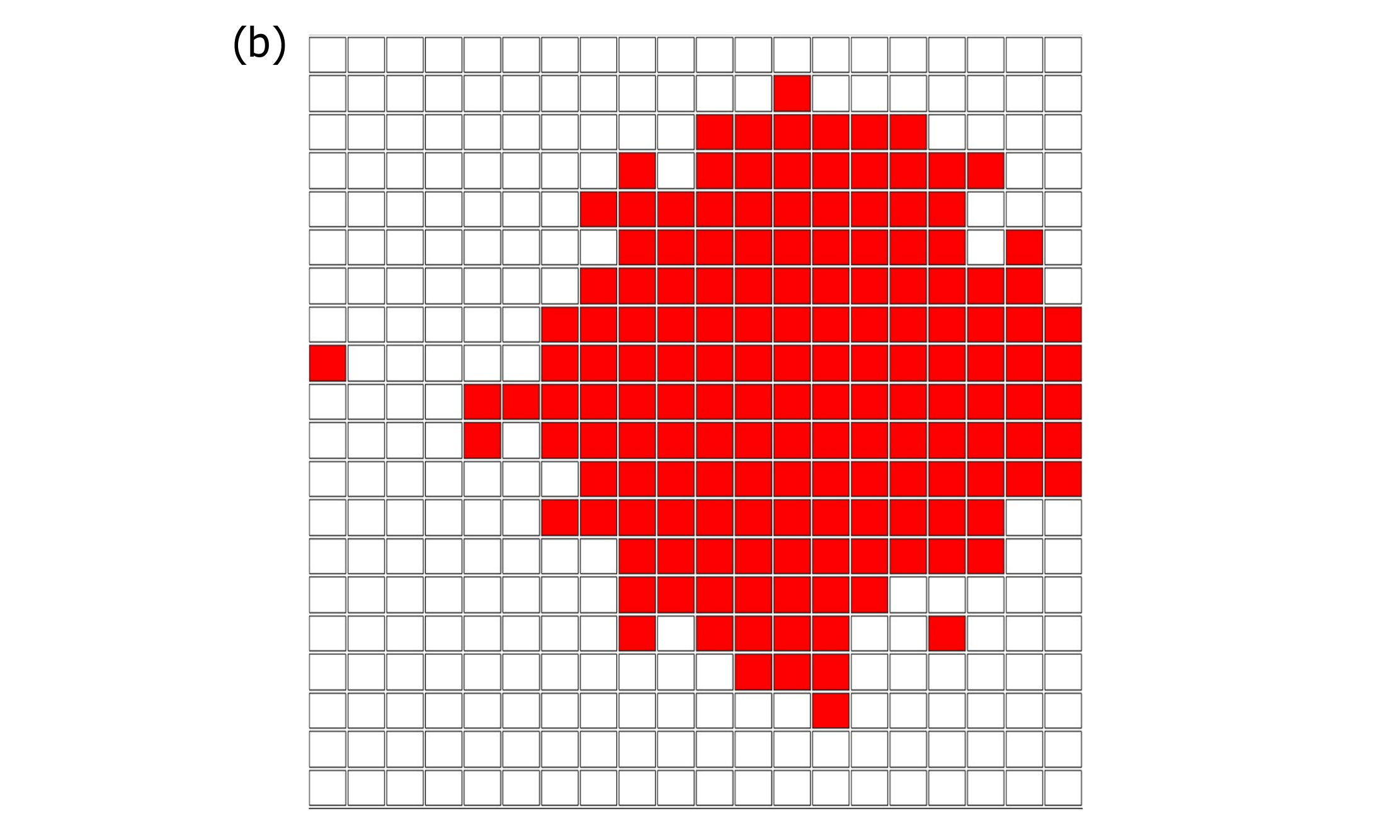}
\includegraphics[ trim=100 0 100 0,clip, width=0.3\linewidth] {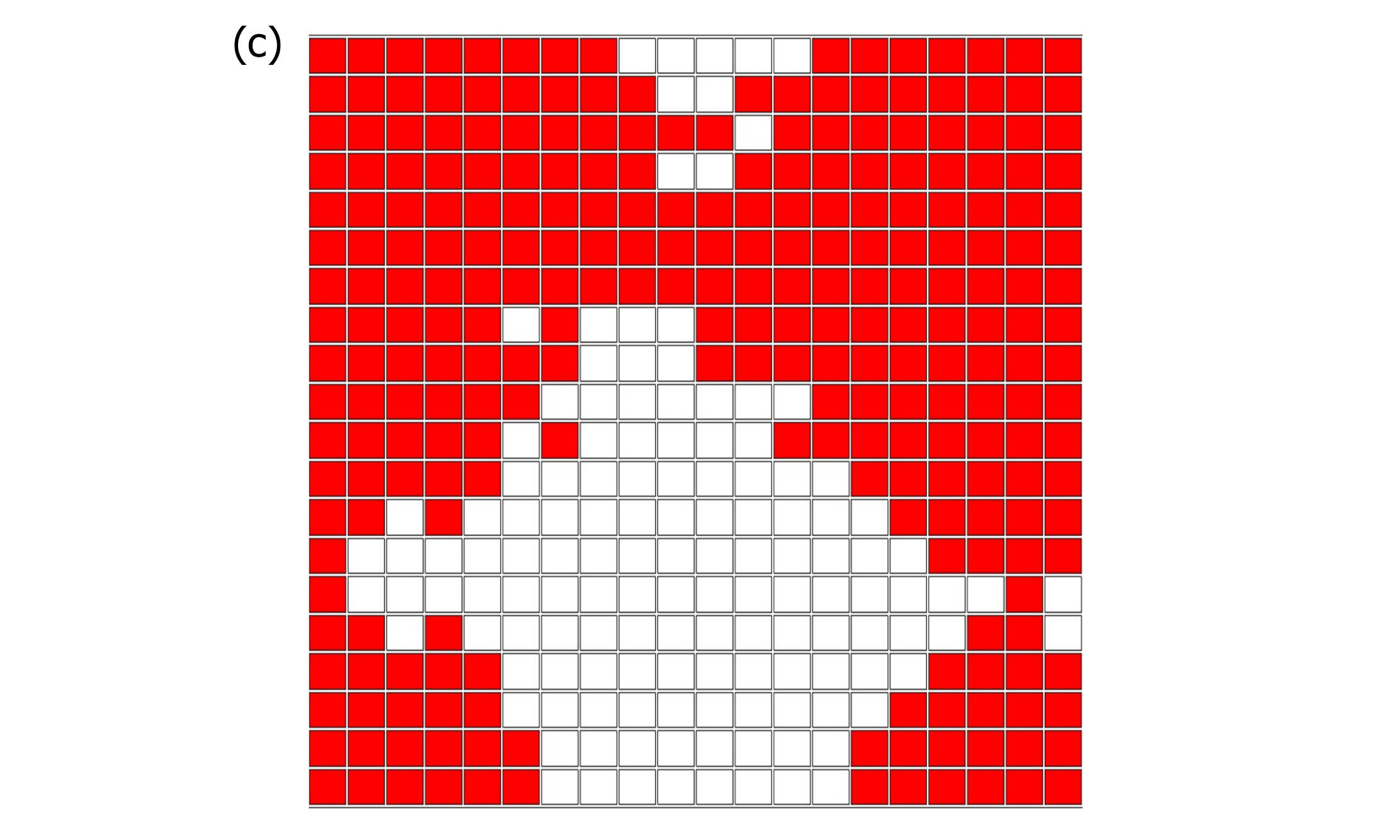}
\caption{
An example of the snapshots of vehicle location on a $20\times 20$ square lattice, with $g=0.6$ and (a) $\rho=0.1$, (b) $\rho=0.4$ and (c) $\rho=0.66$.  Filled sites are occupied by vehicles while unfilled sites are empty. (a)  The system is in the free-flow state and vehicles are roughly evenly distributed on the lattice. (b) The system is in the congested-flow state and most vehicles are stuck in the congested cluster, except vehicles at the cluster periphery. (c) The system is in the congested state, and the congested cluster increases to a size which spans the system.
}
\label{fig_map}
\end{figure*}

Conventionally, the traffic flow $f$ increases with the vehicle density $\rho$ in the \emph{free-flow state}, but $f$ decreases with $\rho$ in the \emph{congested state}~\cite{maerivoit2005_physicsReports}. As we can see in \fig{fig_flow}, in cases with a small path-greediness such as $g=0$ or $g=0.2$, i.e. when routing strategies are very random, the traffic flow $f$ increases to a peak and then decreases which is consistent with the conventional pictures. On the other hand, with a large value of path-greediness such as $g=0.6$ or $g=0.8$, the traffic flow $f$ increases with $\rho$  when $\rho<\rho_c$, implying that the free-flow state emerges when $\rho<\rho_c$. As shown in \fig{fig_map}(a), in the free-flow state, the vehicles can freely flow so that their average speed is high.

\begin{figure*}
\includegraphics[ width=0.45\linewidth] {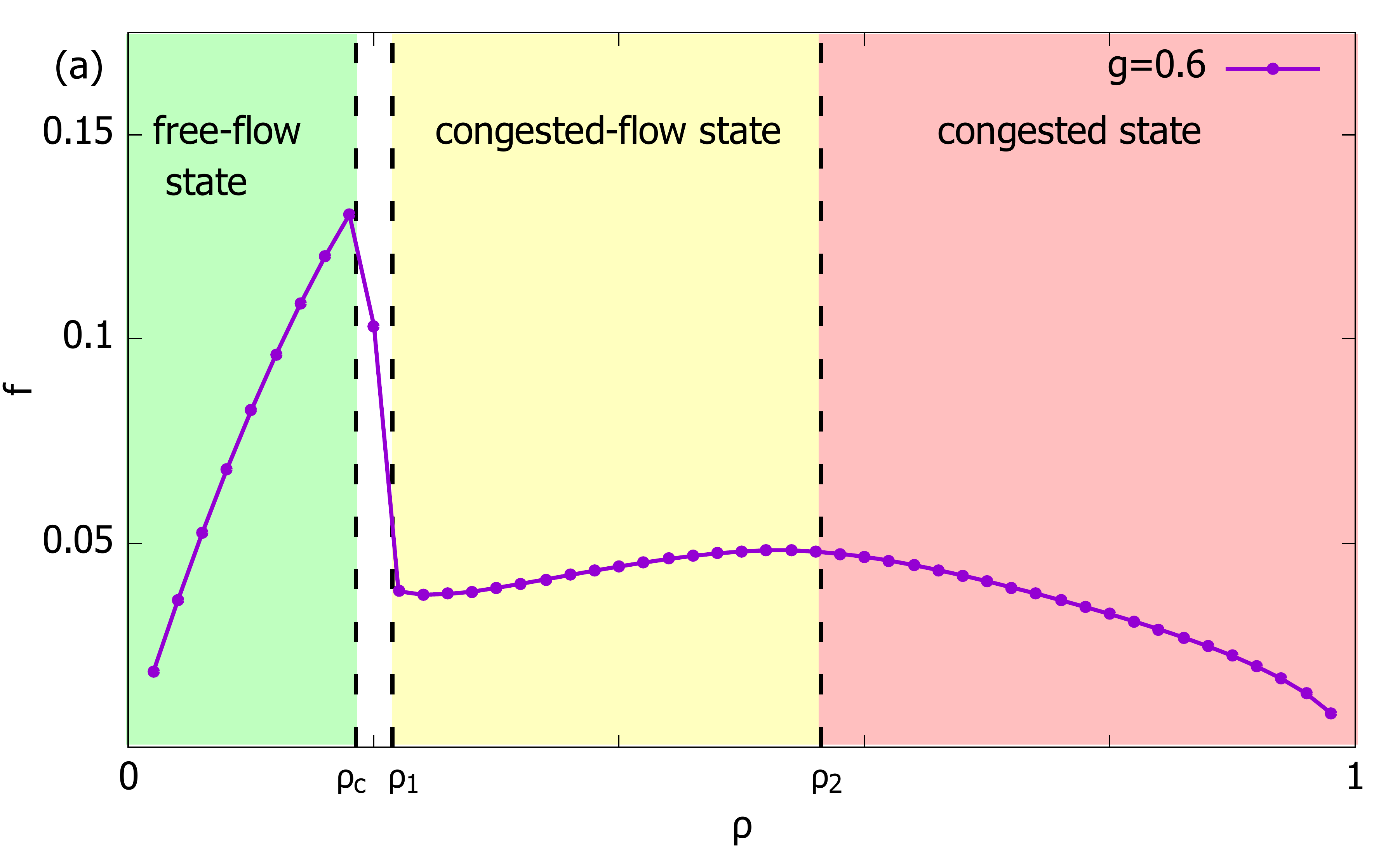}
\includegraphics[ width=0.45\linewidth] {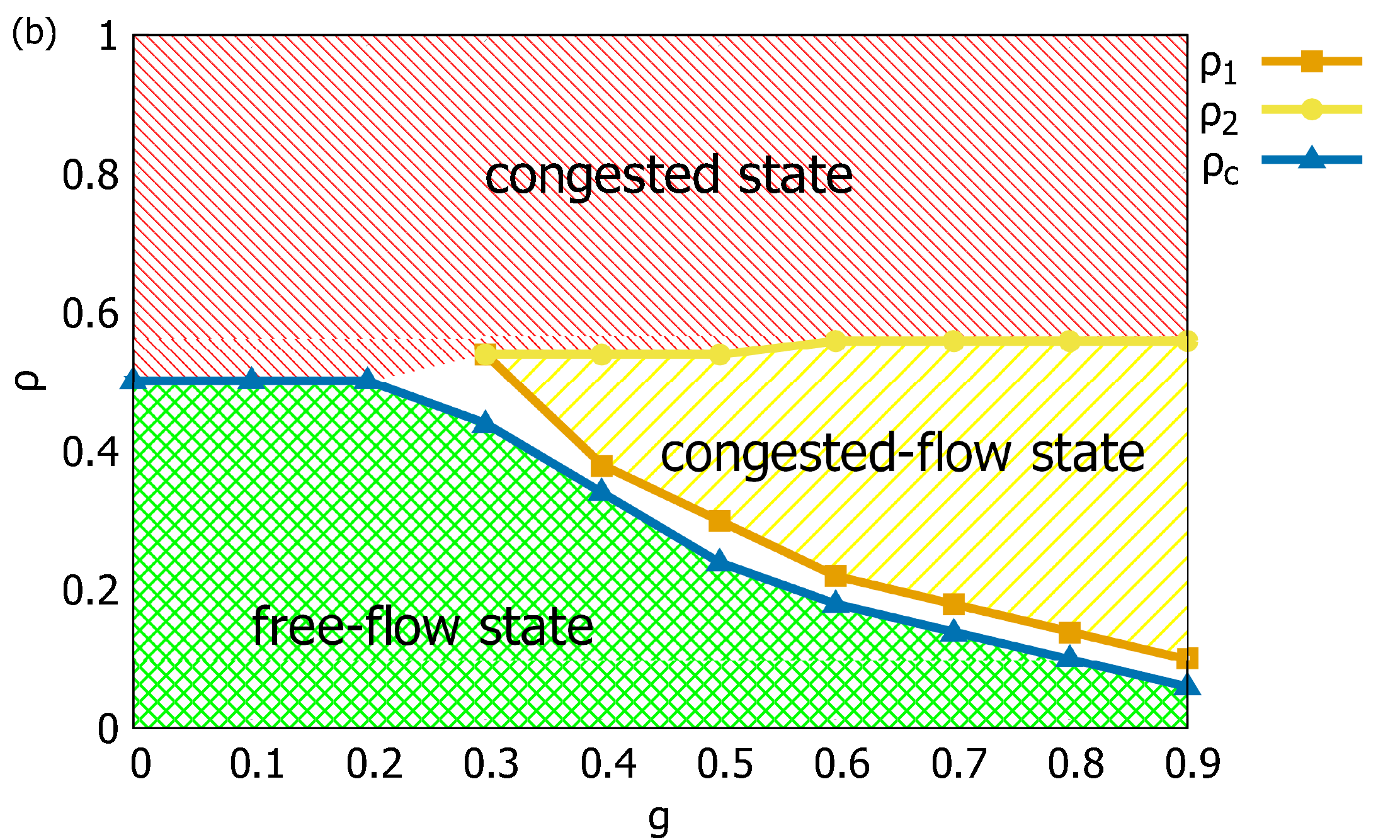}
\caption{
(a) The simulation results of the traffic flow $f$ as a function of $\rho$  with $L=20$ and $g=0.6$. Three states are shown: the free-flow state with $\rho<\rho_c$, the congested-flow state with $\rho_1<\rho<\rho_2$ and the congested state with $\rho<\rho_2$. (b) The phase diagram showing the three states as a function of $\rho$ and $g$.
}
\label{fig_flowPhase}
\end{figure*}

Nevertheless, unlike other conventional studies from which the traffic flow $f$ decreases with $\rho$ beyond the free-flow state, we see in \fig{fig_flow} that for cases with a large path-greediness such as $g=0.6$ or $g=0.8$, the traffic flow slightly increases in a range of values of $\rho$ with $\rho>\rho_c$. To better illustrate the phenomenon, we show specifically the flow $f$ as a function of $\rho$ with $g=0.6$ in \fig{fig_flowPhase}(a). As we can see, $f$ increases slightly in the range of $\rho_1<\rho<\rho_2$, which emerges beyond the free-flow state. Although $\rho_1\neq\rho_c$, we expect $\rho_c$ and $\rho_1$ to coincide when the system size $L$ increases. The results of increasing flow beyond $f_c$ are not observed in other conventional studies, but we remark that the emergence of three phases is also observed in some transportation models~\cite{PhysRevE.84.046110,PhysRevE.76.026105,TIAN2015138}.

We denote the value of $\rho$ at the second peak to be $\rho_2$ as shown in \fig{fig_flowPhase}(a), and call the range $\rho_1<\rho<\rho_2$ with the small increase in flow the \emph{congested-flow state}. In this case, as shown in \fig{fig_map}(b), a congested cluster emerges and block the pathways of many vehicles. The cluster emerges since the destinations of vehicles are randomly assigned, and vehicles with a greedy routing strategy slowly gathered around the cluster if their destinations fall into the clustered region. Vehicles can still freely travel in the uncongested area. Therefore, this congested-flow state is not observed in the conventional one-dimensional cellular automata model, where blocking usually results in complete congestion. At a smaller value of $g$ such as $g\le 0.4$, the congested-flow state is absent as shown in \fig{fig_flow}, and only the free-flow and the congested states are observed. 

Finally, the system emerges into a congested state in which the flow $f$ decreases with $\rho$. For cases with small $g$, the congested state emerges with $\rho>\rho_c$ as shown in \fig{fig_flow}; for cases with large $g$, the congested state emerges with $\rho>\rho_2$ as shown in \fig{fig_flowPhase}(a). As shown in \fig{fig_map}(c), the system in completely congested in both dimensions, leaving only a small central region for vehicle movement. A phase diagram summarizing  the free-flow state, the congested-flow state and the congested state as a function of vehicle density $\rho$ and path-greediness $g$ is shown in \fig{fig_flowPhase}(b).

\begin{figure}

\includegraphics[ width=1\linewidth] {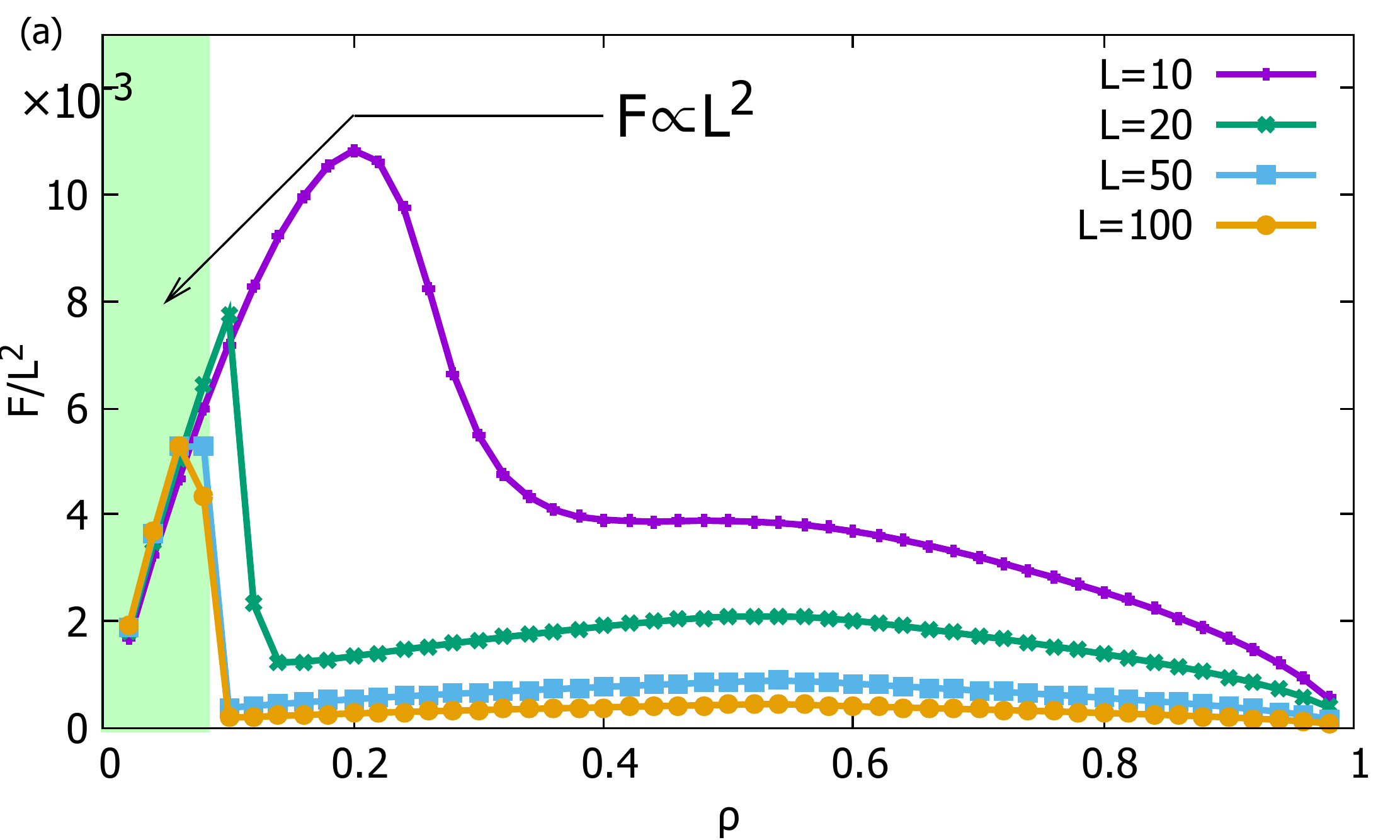}
\includegraphics[ width=1\linewidth] {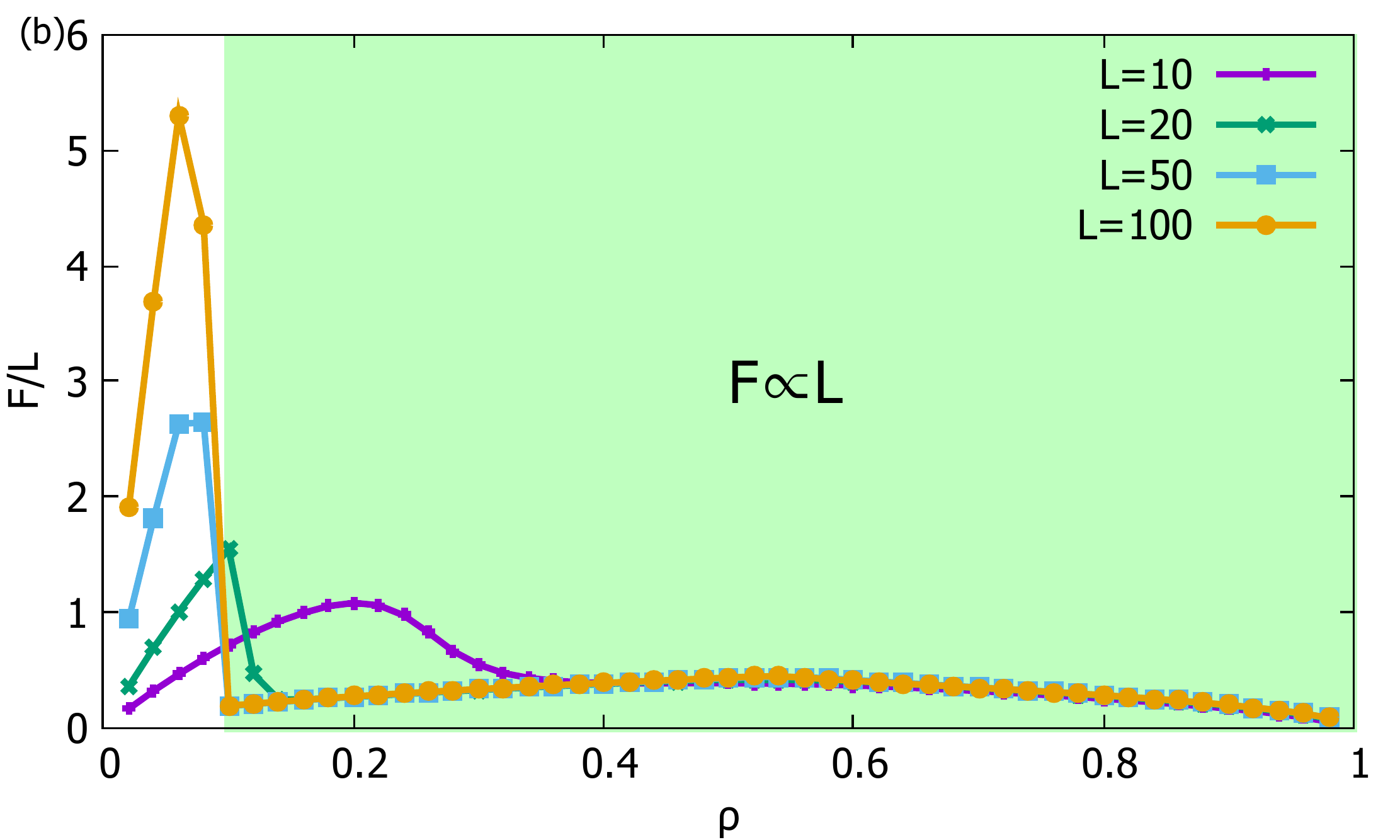}
\caption{
The simulation results of the system traffic flow $F=fL^2$, rescaled as (a) $F/L^2$ and (b) $F/L$, as a function of $\rho$ with $g=0.6$ and various values of $L$. Data collapse is observed in the regime of the free-flow state (i.e. $\rho<\rho_c$) in (a), and the congested-flow and the congested state (i.e. $\rho>\rho_c$) in (b).
}
\label{ratio}
\end{figure}

The emergence of the congested-flow state may be advocated to the reduction of the system dimension from two dimensions to one dimension due to congestion. To reveal this phenomenon, we first define the \emph{system traffic flow} $F=fL^2$, which is the total traffic flow in the whole lattice. We then reveal the effective dimension of the system by examining how the system traffic flow $F$ scales with the system size $L$ in various states.  In \fig{ratio}(a),  the rescaled flow $F/L^2$ obtained from systems of different size collapse in the free-flow state, i.e. with density $\rho<\rho_c$. These results imply that the system is effectively a two-dimensional system in the regime with $\rho<\rho_c$. On the other hand, as shown in \fig{ratio}(b), the results of rescaled flow $F/L$ with different  $L$ collapse in the congested free-flow state and the congested state, i.e. in the regime with $\rho_1<\rho\le 1$, suggesting that the system is reduced to an effectively one-dimensional system. In other words, the system traffic flow $F$ scales as
\begin{align}
F\propto
\begin{cases}
L^2, &\mbox{(Free-flow state)}
\\
L. &\mbox{(Congested-flow and congested state)}
\\
\end{cases}
\end{align}

These results lead to an interpretation of various states in terms of dimensions: the free-flow state is characterized with two-dimensional degree of freedom inherited from the dimension of the original system; it is then reduced to the congested-flow state which is an effectively one-dimensional system.

\subsection{The greediness in routing}

After examining the emergence of various states in the system, we go on to reveal the system dependence on routing strategies.  

\begin{figure*}
\includegraphics[ width=0.49\linewidth] {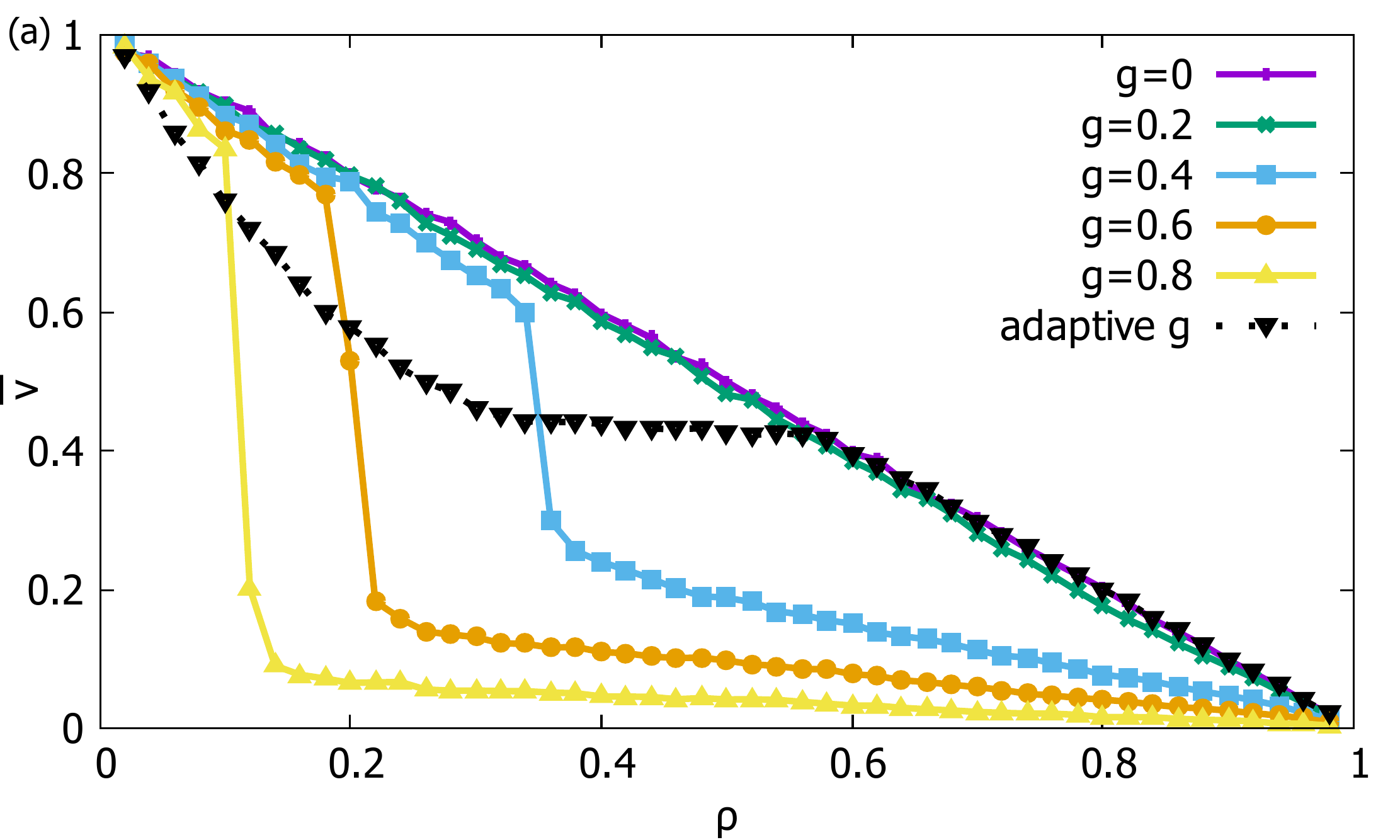}
\includegraphics[ width=0.49\linewidth] {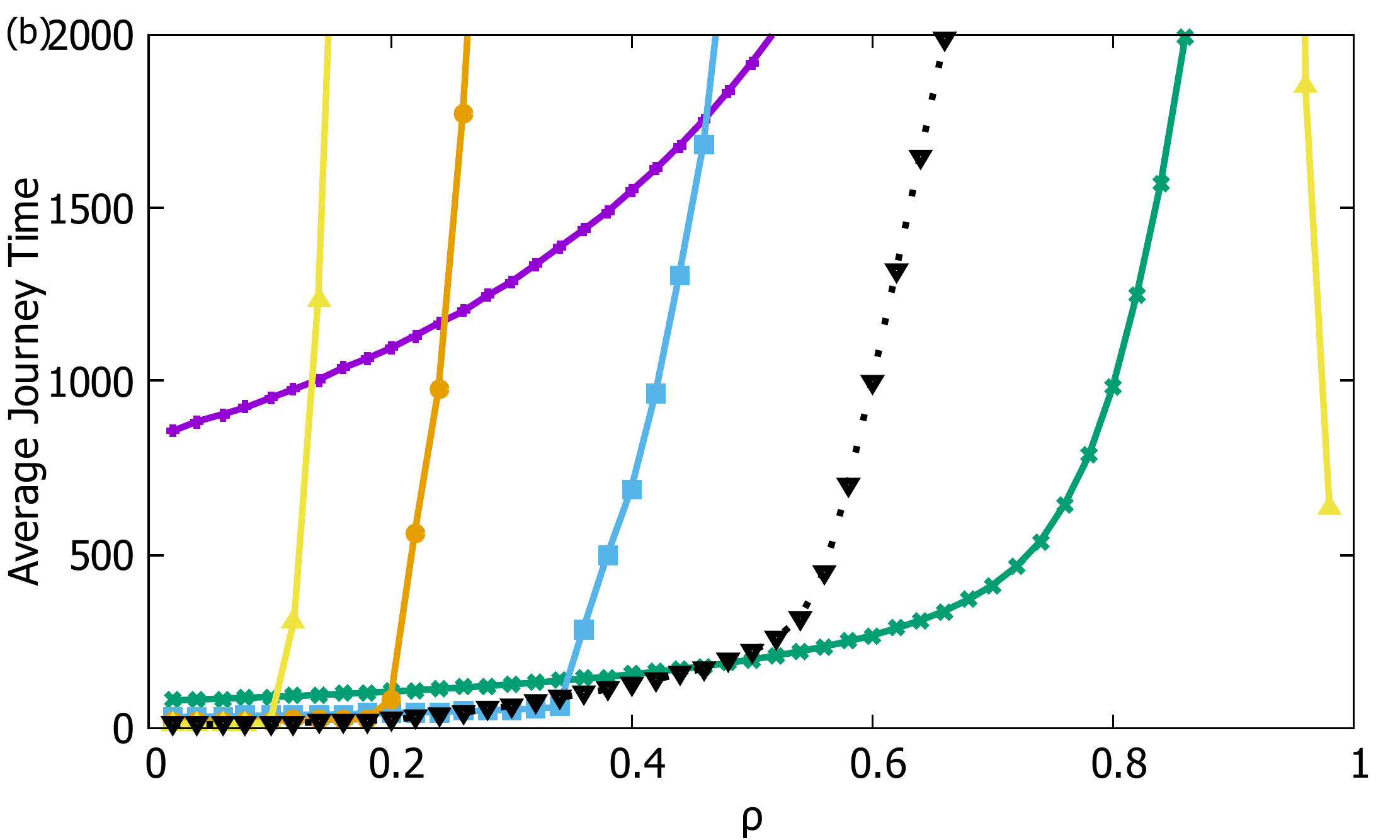}
\includegraphics[ width=0.49\linewidth] {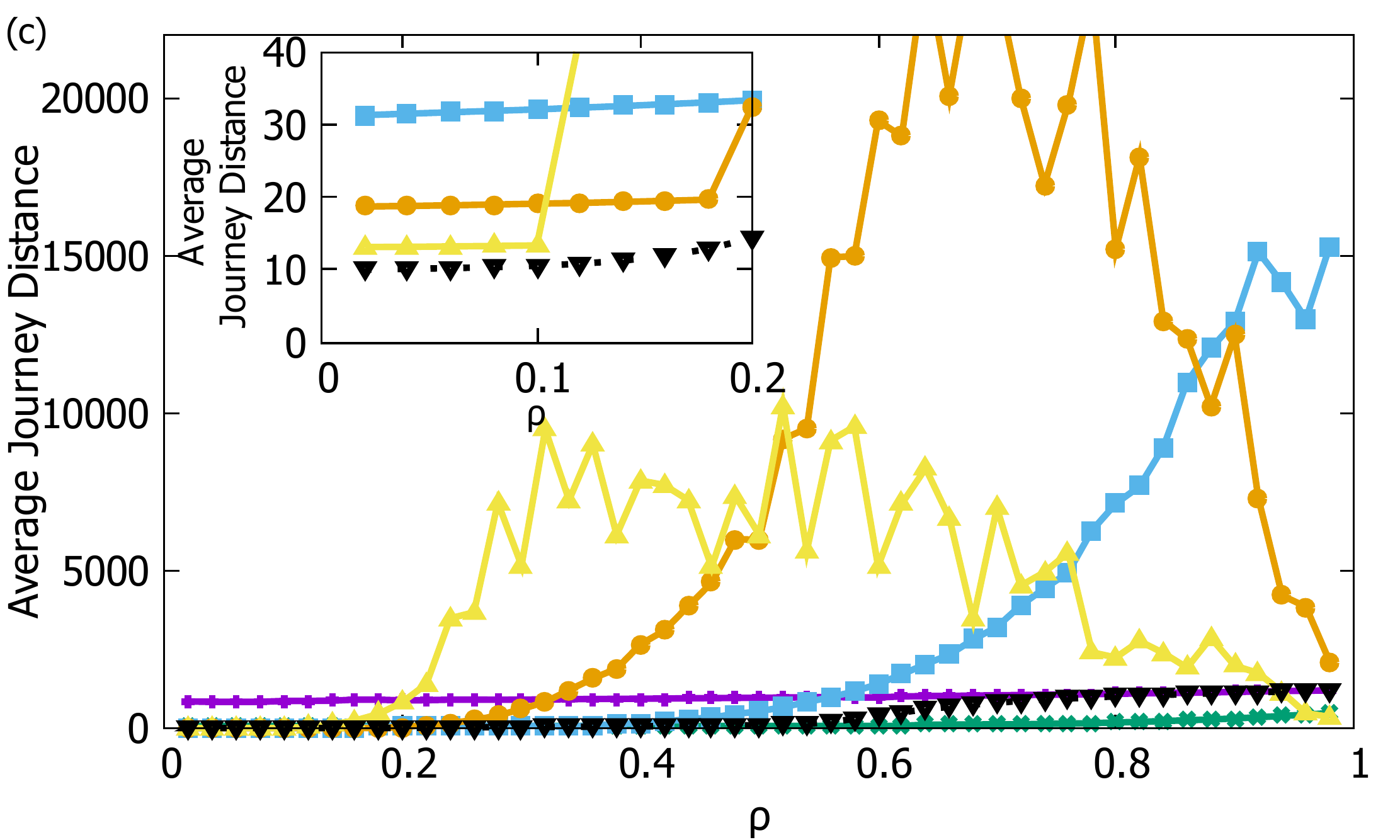}
\includegraphics[ width=0.49\linewidth] {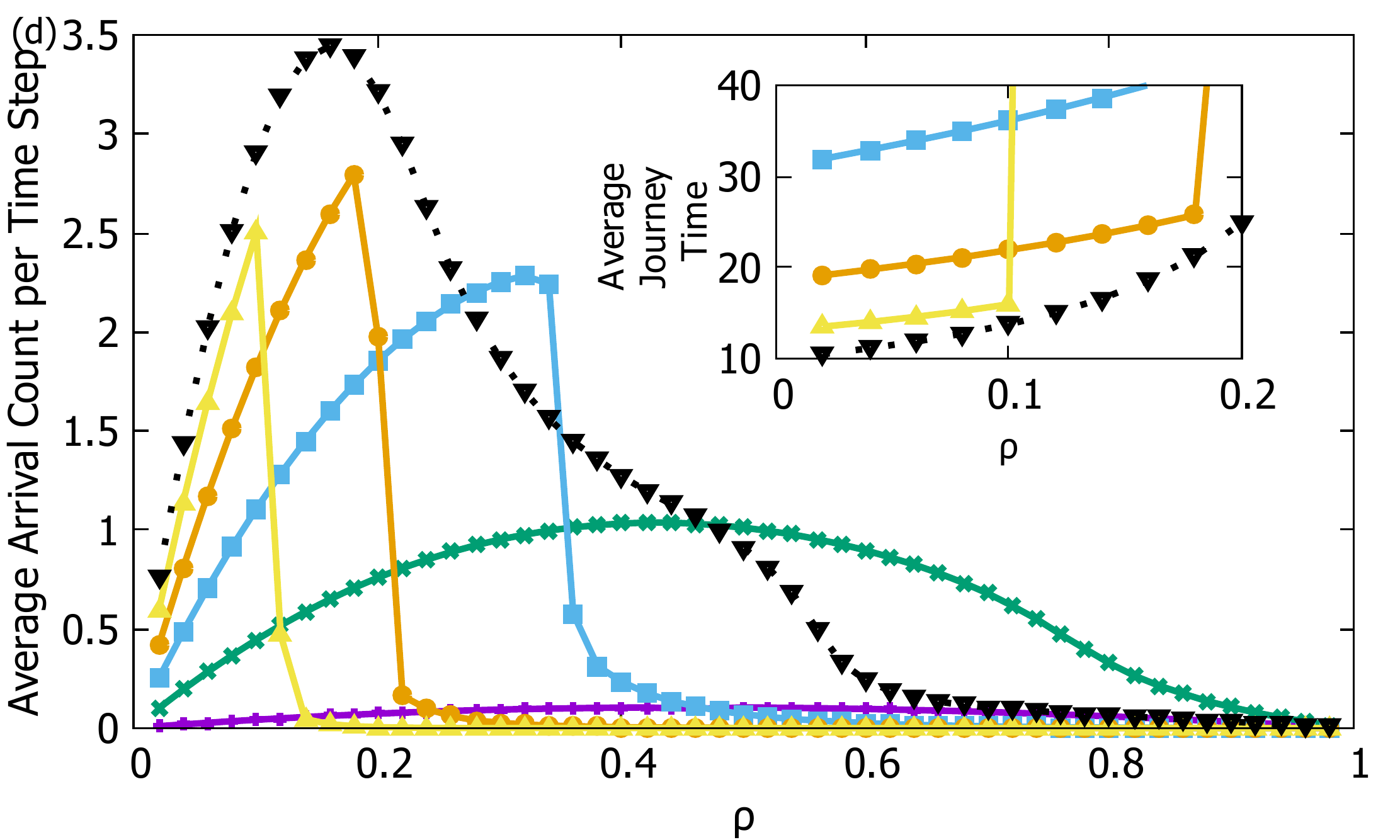}
\caption{
 (a) The average vehicle speed $\bar{v}$, (b) the average journey time  and (c) the average journey distance between the origin and destination of vehicles, as well as (d) the average arrival count per time step as a function of vehicle density $\rho$, for $L=20$ and various values of path-greediness $g$. The corresponding results for the cases with adaptive path-greediness are also shown. The results are obtained with $T=3 \times 10^6$ and an equilibration time $T_e=2.5\times 10^6$, averaged over 1000 instances. Insets: the enlarged plots of the average journey time and distance in the regime with small density $\rho$ are shown in the insets of (d) and (c).
}
\label{fig_constantData}
\end{figure*}

We first examine the dependence of average speed $\bar{v}$ on the path-greediness $g$. As shown in \fig{fig_constantData}(a), the maximum average vehicle speed $\bar{v}=\vmax$ is only observed at the smallest vehicle density $\rho\approx 0$, and $\bar{v}$ decreases linearly even in the free-flow state (i.e. $\rho<\rho_c$) as $\rho$ increases. It is because for each vehicle, its probability to run into another vehicle is $\rho$ in the two-dimensional space even without congestion, resulting in an average speed of $\bar{v}=1-\rho$ in the free-flow state. These results are different from the one-dimensional case of ring, where the average speed $\bar{v}=\vmax$ is constant in the whole free-flow state. As we can see in \fig{fig_constantData}(a), the results in cases with $g=0$, i.e. the cases of random routing, are consistent with the theoretical prediction of $\bar{v}=1-\rho$ for all $\rho$.

For cases with path-greediness $g\ge 0.4$,  $\bar{v}$ decreases abruptly to a small value as $\rho$ increases beyond the threshold density $\rho_c$ as shown in \fig{fig_constantData}(a), indicating the emergence of congestion as we have discussed in Sec.~\ref{sec_phase}. The decrease of average vehicle speed is less for cases with smaller $g$, i.e. when vehicles move more randomly. This is consistent with the observations of larger flow $f$ in cases with smaller $g$ as shown in \fig{fig_flow}, implying that a less greedy strategy results in a larger flow at all values of $\rho$. Nevertheless, since vehicles may move in a random direction, a higher average speed $\bar{v}$ or flow $f$ does not necessary correspond to a higher arrival count, or a short time or distance to the destination as we will discuss below.

Next, we examine the dependence of the arrival count on path-greediness. As shown in \fig{fig_constantData}(d), the arrival count per time step increases with $\rho$ in the free-flow state with $\rho<\rho_c$. It implies that the system can accommodate the commutation of more vehicles as density $\rho$ increases. When $\rho=\rho_c$, the number of arrivals attains its maximum value and the arrival count no longer increases, but instead decreases when there are more vehicles. On the other hand, when path-greediness $g$ increases, i.e. when vehicles have a higher preference to travel via the shortest path to their destinations, we see that (i) the maximum arrival count per time step increases, implying that the system can accommodate more commutations, but (ii) the threshold density $\rho_c$ decreases, implying that congestion emerge at a lower vehicle density.

\subsection{The average journey time and distance}

Other than the average vehicle speed and the arrival count, we examine the most important metrics, i.e. the average journey time and the average journey distance between the origin and the destination of vehicles. As shown in \fig{fig_constantData}(b) and the inset of \fig{fig_constantData}(d), at a specific value of path-greediness $g$, the average journey time is roughly constant when $\rho<\rho_c$. It then abruptly increases at $\rho=\rho_c$ and continues to increase with increasing $\rho$. Similarly as shown in fig \fig{fig_constantData}(c) and its inset, when the system enters the congested state, the average journey distance greatly increases, implying that vehicles are often blocked and have to travel more in random directions in order to move.

We note that both the average journey time and journey distance increase abruptly in the congested state. The reason is that in the congested state, vehicles are stopped for a large number of steps before spaces are spared for them to move. In particular, the average journey distance increases abruptly when congestion occur, as vehicles need to travel for many unwanted or unnecessary steps. If the destinations of vehicles are in the congested cluster, they have to move around the cluster in order to find a route into the cluster. A lot of movements are used for attempts and to retrieve from their unsuccessful trials. When the density of vehicles is well beyond the threshold density, i.e. $\rho\gg \rho_c$, the congested cluster becomes larger, and unsuccessful attempts are more frequent so that the average journey distance increases drastically.

Next, we examine how the average journey time and distance are affected by the path-greediness. As we can see in \fig{fig_constantData}(a), the average vehicle speed $\bar{v}$ is higher in cases with small $g$, which indicates that the system is less congested. Nevertheless, it does not necessarily imply that vehicles have a shorter journey time, since the movements of vehicles are random at small $g$. This can be shown by the average journey time in the inset of \fig{fig_constantData}(d), when the vehicle density is small, the smaller the value of $g$, the more random the path, and the longer the time it takes for the vehicles to arrive at their destinations. Remarkably, these expected results are true only at small vehicle density $\rho$. In contrast to our common belief, as shown by \fig{fig_constantData}(b), at large $\rho$ the average journey time is shorter with smaller values of $g$; this suggests that a less greedy routing strategy at large vehicle density $\rho$ would lead to a short average journey time. 

Similarly, as we can see in the inset of \fig{fig_constantData}(c),  the greedy cases (i.e. large $g$) have a shorter average journey distance than the less greedy cases (i.e. small $g$) in the free-flow state, but the opposite is true in the congested state at large $\rho$ as shown by \fig{fig_constantData}(c). Similar to our observations on the average journey time, the results on the average journey distance imply that a less greedy (or more random) routing strategy is beneficial at large vehicle density $\rho$, i.e. in highly congested states.

Furthermore, from the results of \fig{fig_constantData}(d), there exist a value of path-greediness $g$ for each density $\rho$ which maximizes the arrival count. One interesting interpretation of the results in \fig{fig_constantData}(d) is that when the vehicle density $\rho$ increases beyond $\rho_c$, one can keep the system in the free-flow state by decreasing the path-greediness of the system. It means that if some vehicles can travel via longer paths, the whole system benefits as the average journey time and distance in the free-flow state are always shorter than those in the congested state.

\subsection{Vehicles with adaptive path-greediness}

To further reveal the impact of routing strategies on the macroscopic behavior of transportation networks, finally we examine a case in which individuals can self-adjust their own path-greediness. In this scenario, each vehicle $i$ is assigned the same initial value of $g_i$. When simulations start, vehicle $i$ first follows the initial $g_i$ in movements. Then, $g_i$ increases by $\Delta g$ when vehicle $i$ has moved successfully for $P$ consecutive steps without being blocked by the other vehicles. This corresponds to the case that vehicle $i$ perceives the network as in the free-flow state and a higher greediness would shorten the time to the destination. On the other hand, $g_i$ decreases by $\Delta g$ if vehicle $i$ has been stopped for $P$ consecutive steps, as it prefers a longer route when the shorter routes are congested. For simplicity, we set $\Delta g = 0.04 $ and $P=3$ in simulations. In this adaptive case, we observe that the equilibrium state of the system is independent of the initial value of $g$.

For comparison, we call the original case  with a universal constant $g$ the \emph{controlled case}, while the case with an individual adaptive $g$ the \emph{adaptive case}. In general, we found that the adaptive case outperforms the controlled case in the free-flow state with $\rho<\rho_c$, but not in the congested state with $\rho<\rho_c$ . For instance, as shown in the insets of \fig{fig_constantData}(c) and (d), the adaptive case has a lower average journey time and distance than those in the controlled case. In other words, vehicles are able to adjust their $g$ in order to shorten their journey time and distance in the free-flow state. Furthermore, in \fig{fig_constantData}(d), the arrival count in the adaptive case is always higher than that of the controlled case when the vehicle density is low. Nevertheless, with large values of $\rho$, the adaptive case has a larger average journey time and distance as well as a smaller arrival count compared with those of the controlled case, implying that the adaptive case does not perform as well as the controlled case in the congested state.

\begin{figure}
\includegraphics[ width=1\linewidth] {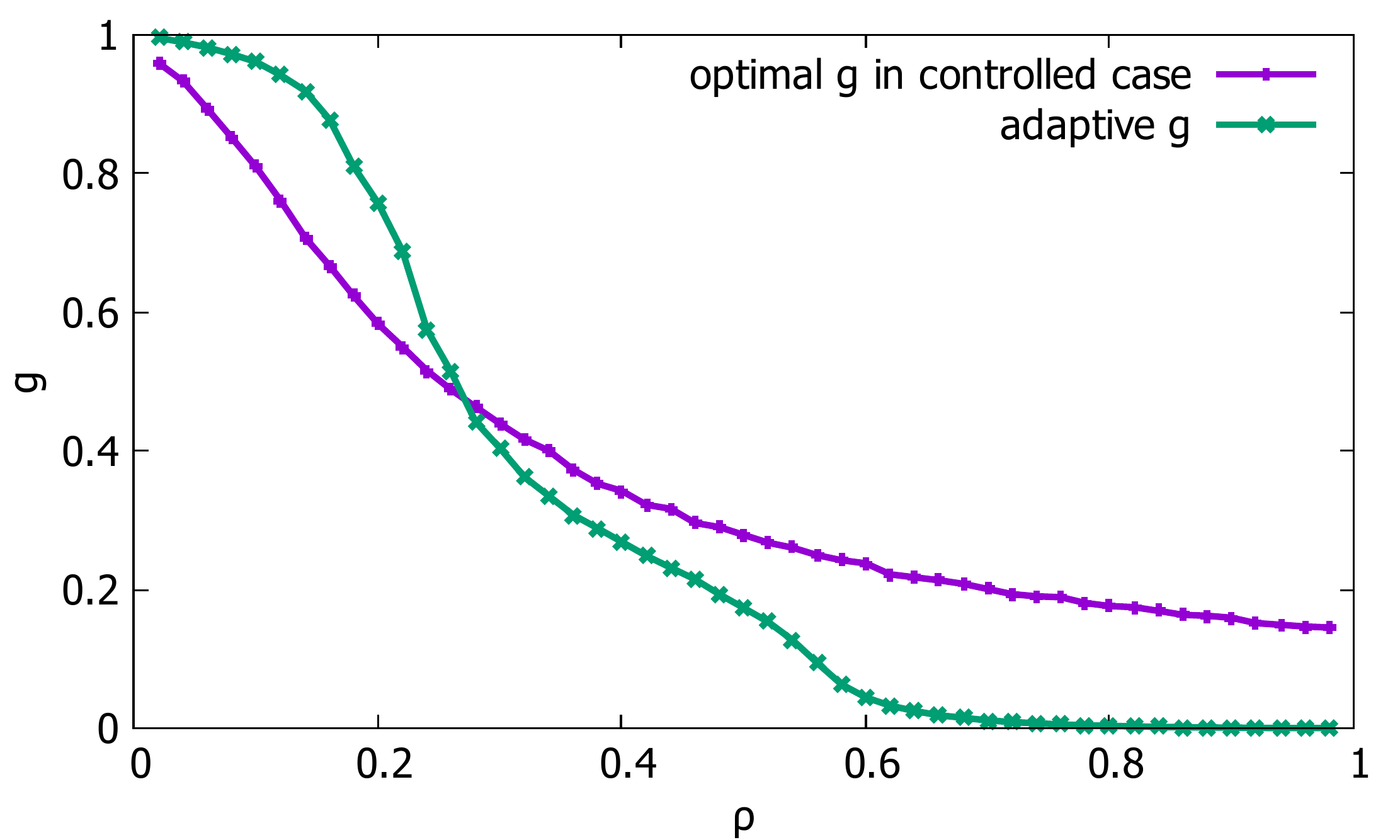}
\caption{
The simulation results of the optimal value of $g$ identified in the controlled case for each density $\rho$, compared with the average value of $g$ in the adaptive case, on $20\times 20$ square lattices with $T=3 \times 10^6$ and an equilibration time $T_e=2.5\times 10^6$, averaged over 1000 instances.
}
\label{fig_constantVSchanging}
\end{figure}

To better compare the controlled and the adaptive cases, for each value of vehicle density, we identify the optimal value of $g$ in the controlled case which maximizes the arrival count  in \fig{fig_constantData}(d). In \fig{fig_constantVSchanging}, we compare the optimal value of $g$ in the controlled case with the average $g$ in the adaptive case as a function of $\rho$. As we have discussed above, the adaptive case outperforms the controlled case in terms of the arrival count in the free-flow state; \fig{fig_constantVSchanging} shows that vehicles self-adapt to a higher $g$ value in the adaptive case than that in the controlled case with small $\rho$, implying that a distribution of path-greediness among vehicles are beneficial in the free-flow state, since those vehicles in the less-crowded region can adopt a higher $g$ to shorten their journey time and distance. 

Nevertheless, the opposite is true in the congested state. In this case, the controlled case outperforms the adaptive case, and the controlled values of $g$ are higher than those in the adaptive case. These results suggest that in the congested case, a controlled environment leads to a better coordination among vehicles, and hence a higher arrival count than that of the adaptive case.

\section{Conclusion}

In this paper, we introduced a model of transportation networks which is based on two-dimensional cellular automata. In the model, individual vehicles travel to their respective destinations with a routing strategy characterized by a single parameter called path-greediness, which corresponds to their tendency to travel via the shortest paths to the destinations. 

We found that in cases with a high path-greediness, three states including the free-flow state, the congested-flow state and the congested state emerge. In the congested-flow state, the traffic flow increases with the vehicle density even in the presence of congestion. One may advocate the emergence of congested-flow state with the reduction of the effective dimension of the system, as shown by our rescaled results. Moreover, a high tendency to travel to the destination via the short path does not necessarily increase the total arrival count or shorten the average journey time; contrary to our common belief, a less greedy routing strategy may be beneficial in the congested state. Finally, by comparing cases where vehicles self-adjust their own path-greediness to cases with a globally controlled path-greediness, we found that adaptivity is only beneficial in the free-flow state, but not in the congested state. These results imply that coordination among vehicles may be induced by a universal path-greediness, which is beneficial for suppressing congestion.

In summary, our results imply that a small tendency to travel via the shortest path and a centralized control in routing strategies are beneficial to transportation networks with congestion, which are in contrast to our common belief. Our results point to the importance in the relation between routing strategies, traffic control and traffic congestion, which further shed light on the mitigation of congestion via route coordination. These insights are also relevant to rapidly developing technologies with enhance control on individual vehicles, including self-driving cars and intelligent transportation system.

\begin{acknowledgments}
The work is fully supported by the Research Grants Council of Hong Kong Special Administrative Region, China (Project No. EdUHK 28300215, EdUHK 18304316 and EdUHK 18301217).
\end{acknowledgments}

\appendix
\section{The cases with non-unity speed limit}
\label{sec_vmax}

Here we consider the cases with speed limit $\vmax>1$. In this case, vehicles are able to hop more than once in a single time step. We thus divide a single time step into $\vmax$ \emph{sub-steps} labelled by $n=0, 1,\dots,\vmax$ and define the sub-step time interval to be $\delta t = 1/\vmax$. We randomly pick a vehicle $i$ at time $t$ and denote the coordinates of vehicle $i$ at the $n$-th sub-step to be $(\xitn, \yitn)$; when $n=0$, the vehicle is in coordinate $(\xit,\yit)$ and when $n=\vmax$, the vehicle is in coordinate $(\xito,\yito)$. The coordinate of vehicle $i$ at the $(n+1)$-th sub-step is updated by
\begin{align}
\label{eq_substep}
&\tuple{\xitno}{\yitno} 
\nonumber\\
&\quad\quad = \tuple{\xitn}{\yitn} + \eta_i^{t+n\delta t}\tuple{\Delta \xitn}{\Delta \yitn}.
\end{align}
where $\left(\Delta \xitn, \Delta \yitn \right)$ corresponds to the intended movement of vehicle $i$ in the $n$-th sub-step at time $t$, and $\eta_i^{t+n\delta t}=0$ if the site of the next intended movement is occupied by another vehicle, and otherwise $\eta_i^{t+n\delta t}=1$. In other words, $\eta_i^{t+n\delta t}$ is given by
\begin{align}
\eta_i^{t+n\delta t} =
\begin{cases}
0, &\hspace{-0.2cm}\mbox{if $\tuple{\xitn}{\yitn}\!+\!\tuple{\Delta \xitn}{\Delta \yitn}\!$}
\nonumber\\
&\mbox{\quad\quad\quad\quad\quad\quad\quad\quad\quad\quad$=\!\tuple{\xjt}{\yjt}, \exists j$}
\\
1, &\hspace{-0.2cm}\mbox{otherwise}
\end{cases}
\end{align}
In this case, the movement of vehicle $i$ from coordinate $(\xit,\yit)$ at time $t$ to $(\xito,\yito)$ at time $t+1$ is composed of $\vmax$ single movements, as we can see by summing \req{eq_substep} from $n=0$ to $n=\vmax-1$, i.e.
\begin{align}
\tuple{\xito}{\yito}=\tuple{\xit}{\yit} + \sum_{n=1}^{\vmax}\eta_i^{t+n\delta t}\tuple{\Delta \xitn}{\Delta \yitn},
\end{align}
where each movement is denoted as $\eta_i^{t+n\delta t}(\Delta \xitn, \Delta \yitn)$ with $n=1,\dots,\vmax$.

As discussed in Sec.~\ref{sec_model}, we express the intended movement $(\Delta \xitn, \Delta \yitn)$ in terms of path-greediness $g$.  When  vehicle $i$ has not yet arrived at the $x$- nor $y$-coordinate of its destination (i.e. $\xitn\neq\dix$ and $\yitn\neq\diy$), we draw $(\Delta \xitn, \Delta \yitn)$ according to
\begin{align}
\label{eq_random1}
&\left(\Delta \xitn, \Delta \yitn\right)=
\nonumber\\
&\quad\quad\quad\quad\quad
\begin{cases}
\tuple{0}{\Delta \tilde{y}}, &\mbox{with a probability $\frac{1+g}{4}$},
\\
\tuple{\Delta \tilde{x}}{0}, &\mbox{with a probability $\frac{1+g}{4}$},
\\
\tuple{0}{-\Delta \tilde{y}}, &\mbox{with a probability $\frac{1-g}{4}$},
\\
\tuple{-\Delta \tilde{x}}{0}, &\mbox{with a probability $\frac{1-g}{4}$},
\\
\end{cases}
\nonumber\\
\end{align}
where $\Delta \tilde{y}$ and  $\Delta \tilde{x}$ are the greedy directions
\begin{align}
\Delta \tilde{y}  &= \sgn{Y_i-\yitn} \sgn{\frac{L}{2}-|Y_i-\yitn|},
\\
\Delta \tilde{x}  &= \sgn{X_i-\xitn} \sgn{\frac{L}{2}-|X_i-\xitn|}.
\end{align}
On the other hand, when vehicle $i$ has arrived at either the $x$- or $y$-coordinate of its destination (i.e. $\xitn=X_i$ and $\yitn\neq Y_i$, or $\xitn\neq X_i$ and $\yitn= Y_i$), we assume that the vehicle moves in a direction towards the destination with a probability $(1+3g)/4$, such that if $\xitn= X_i$, then
\begin{align}
\label{eq_random1}
&\left(\Delta \xitn, \Delta \yitn\right)=
\nonumber\\
&\quad\quad\quad\quad\quad
\begin{cases}
\tuple{0}{\Delta \tilde{y}},&\mbox{with a probability $\frac{1+3g}{4}$},
\\
\tuple{\Delta \tilde{x}}{0},&\mbox{with a probability $\frac{1-g}{4}$},
\\
\tuple{0}{-\Delta \tilde{y}},&\mbox{with a probability $\frac{1-g}{4}$},
\\
\tuple{-\Delta \tilde{x}}{0}, &\mbox{with a probability $\frac{1-g}{4}$}.
\\
\end{cases}
\end{align}
On the other hand, if $y_n= Y_i$, then
\begin{align}
\label{eq_random1}
&\left(\Delta \xitn, \Delta \yitn\right)=
\nonumber\\
&\quad\quad\quad\quad\quad
\begin{cases}
\tuple{\Delta \tilde{x}}{0},&\mbox{with a probability $\frac{1+3g}{4}$},
\\
\tuple{0}{\Delta \tilde{y}},&\mbox{with a probability $\frac{1-g}{4}$},
\\
\tuple{-\Delta \tilde{x}}{0},&\mbox{with a probability $\frac{1-g}{4}$},
\\
\tuple{0}{-\Delta \tilde{y}}, &\mbox{with a probability $\frac{1-g}{4}$},
\\
\end{cases}
\end{align}
In this case with $\vmax$ sub-steps in each time step, the traffic flow $f$ is given by
\begin{align}
f = \frac{1}{(T-T_e)L^2} \sum_{t=T_e}^{T} \sum_{i=1}^{N} \sum_{n=1}^{\vmax}\eta_i^{t+n\delta t}.
\end{align}




\begin{thebibliography}{10}

\bibitem{PhysRevLett.108.208701}
C.~H. Yeung and D.~Saad,
\newblock Phys. Rev. Lett. {\bf 108}, 208701 (2012).

\bibitem{Yeung13717}
C.~H. Yeung, D.~Saad and K.~Y.~M. Wong,
\newblock Proceedings of the National Academy of Sciences {\bf 110}, 13717
  (2013).

\bibitem{youn2008price}
H.~Youn, M.~T. Gastner and H.~Jeong,
\newblock Phys. Rev. Lett. {\bf 101}, 128701 (2008).

\bibitem{NSModel}
K.~Nagel and M.~Schreckenberg,
\newblock Journal de Physique I {\bf 2}, 2221 (1992).

\bibitem{PhysRevE.53.4655}
K.~Nagel,
\newblock Phys. Rev. E {\bf 53}, 4655 (1996).

\bibitem{CHOWDHURY2000199}
D.~Chowdhury, L.~Santen and A.~Schadschneider,
\newblock Physics Reports {\bf 329}, 199  (2000).

\bibitem{PhysRevE.57.2441}
P.~M. Simon and H.~A. Gutowitz,
\newblock Phys. Rev. E {\bf 57}, 2441 (1998).

\bibitem{DAGANZO2006396}
C.~F. Daganzo,
\newblock Transport. Res. B: Meth. {\bf 40}, 396  (2006).

\bibitem{PhysRevE.84.046110}
B.~S. Kerner, S.~L. Klenov and M.~Schreckenberg,
\newblock Phys. Rev. E {\bf 84}, 046110 (2011).

\bibitem{PhysRevE.51.2939}
M.~Schreckenberg, A.~Schadschneider, K.~Nagel and N.~Ito,
\newblock Phys. Rev. E {\bf 51}, 2939 (1995).

\bibitem{JABARI2012156}
S.~E. Jabari and H.~X. Liu,
\newblock Transport. Res. B: Meth. {\bf 46}, 156  (2012).

\bibitem{SUMALEE2011507}
A.~Sumalee, R.~Zhong, T.~Pan and W.~Szeto,
\newblock Transport. Res. B: Meth. {\bf 45}, 507  (2011).

\bibitem{maerivoit2005_physicsReports}
S.~Maerivoet and B.~D. Moor,
\newblock Physics Reports {\bf 419}, 1  (2005).

\bibitem{VANLINT201863}
J.~van Lint and S.~Calvert,
\newblock Transport. Res. B: Meth. {\bf 117}, 63  (2018).

\bibitem{RAMEZANI20151}
M.~Ramezani, J.~Haddad and N.~Geroliminis,
\newblock Transport. Res. B: Meth. {\bf 74}, 1  (2015).

\bibitem{ZHONG2017292}
R.~Zhong {\em et~al.},
\newblock Transportation Research Part C: Emerging Technologies {\bf 85}, 292
  (2017).

\bibitem{GU20181}
Z.~Gu, S.~Shafiei, Z.~Liu and M.~Saberi,
\newblock Transportation Research Part C: Emerging Technologies {\bf 95}, 1
  (2018).

\bibitem{XU201882}
G.~Xu, H.~Yang, W.~Liu and F.~Shi,
\newblock Transportation Research Part C: Emerging Technologies {\bf 95}, 82
  (2018).

\bibitem{ZHANG2018190}
J.~Zhang, R.~Lindsey and H.~Yang,
\newblock Transport. Res. B: Meth. {\bf 117}, 190
  (2018).

\bibitem{HUANG2017169}
Y.~Huang, L.~Zhao, T.~V. Woensel and J.-P. Gross,
\newblock Transport. Res. B: Meth. {\bf 95}, 169  (2017).

\bibitem{PhysRevE.82.066107}
M.~Kanai,
\newblock Phys. Rev. E {\bf 82}, 066107 (2010).

\bibitem{LE2017539}
T.~Le {\em et~al.},
\newblock Transport. Res. B: Meth. {\bf 105}, 539
  (2017).

\bibitem{ZHAO201887}
C.-L. Zhao and L.~Leclercq,
\newblock Transport. Res. B: Meth. {\bf 117}, 87  (2018).

\bibitem{SUBRAMANYAM2018296}
A.~Subramanyam, A.~Wang and C.~E. Gounaris,
\newblock Transport. Res. B: Meth. {\bf 117}, 296
  (2018).

\bibitem{KOSTER2018137}
P.~Koster, E.~Verhoef, S.~Shepherd and D.~Watling,
\newblock Transport. Res. B: Meth. {\bf 117}, 137
  (2018).

\bibitem{Silva5643}
R.~Silva, S.~M. Kang and E.~M. Airoldi,
\newblock Proceedings of the National Academy of Sciences {\bf 112}, 5643
  (2015).

\bibitem{SIQUEIRA20161}
A.~F. Siqueira, C.~J. Peixoto, C.~Wu and W.-L. Qian,
\newblock Transport. Res. B: Meth. {\bf 87}, 1  (2016).

\bibitem{PhysRevE.76.026105}
K.~Gao, R.~Jiang, S.-X. Hu, B.-H. Wang and Q.-S. Wu,
\newblock Phys. Rev. E {\bf 76}, 026105 (2007).

\bibitem{TIAN2015138}
J.~Tian, M.~Treiber, S.~Ma, B.~Jia and W.~Zhang,
\newblock Transport. Res. B: Meth. {\bf 71}, 138  (2015).

\end{thebibliography}

\end{document}